\documentclass[]{interact}

\usepackage{epstopdf}% To incorporate .eps illustrations using PDFLaTeX, etc.
\usepackage[caption=false]{subfig}% Support for small, `sub' figures and tables
\usepackage[numbers,sort&compress]{natbib}% Citation support using natbib.sty
\bibpunct[, ]{[}{]}{,}{n}{,}{,}% Citation support using natbib.sty
\renewcommand\bibfont{\fontsize{10}{12}\selectfont}% Bibliography support using natbib.sty

\theoremstyle{plain}% Theorem-like structures provided by amsthm.sty

\theoremstyle{definition}

\theoremstyle{remark}

\newcommand{\Sig}{\mbox{\boldmath$\Sigma$\unboldmath}}

\newcommand{\y}{\mbox{\boldmath$y$\unboldmath}}
\newcommand{\bbeta}{\mbox{\boldmath$\beta$\unboldmath}}

\newcommand{\bx}{\mbox{\boldmath$x$\unboldmath}}

\newcommand{\btheta}{\mbox{\boldmath$\theta$\unboldmath}}

\newcommand{\zero}{\mbox{\boldmath$0$\unboldmath}}

\newcommand{\bphi}{\mbox{\boldmath$\phi$\unboldmath}}

\newcommand{\bv}{\mbox{\boldmath$v$\unboldmath}}

\newcommand{\bpsi}{\mbox{\boldmath$\psi$\unboldmath}}
\newcommand{\bvarsigma}{\mbox{\boldmath$\varsigma$\unboldmath}}

\begin{document}

\articletype{ARTICLE TEMPLATE}% Specify the article type or omit as appropriate

\title{Statistical modeling of groundwater quality assessment in Iran using a flexible Poisson likelihood}

\author{
\name{Mahsa Nadifar$^*$, Hossein Baghishani$^*$, Afshin Fallah$^{\dagger}$, H{\aa}vard Rue$^{\ddagger}$ \\ 
$^*$ Department of Statistics, Faculty of Mathematical Sciences, Shahrood University of Technology, Iran\\ $^{\dagger}$ Department of Statistics, Imam Khomeini International University, Qazvin, Iran \\ $^{\ddagger}$ CEMSE Division, King Abdullah University of Science and Technology, Thuwal, Saudi Arabia}
}

\maketitle

\begin{abstract}
Assessing water quality and recognizing its associated risks to human health and the broader environment is undoubtedly essential. Groundwater is widely used to supply water for drinking, industry, and agriculture purposes. The groundwater quality measurements vary for different climates and various human behaviors, and consequently, their spatial variability can be substantial. In this paper, we aim to analyze a groundwater dataset from the Golestan province, Iran, for November 2003 to November 2013. Our target response variable to monitor the quality of groundwater is the number of counts that the quality of water is good for a drink. Hence, we are facing spatial count data. Due to the ubiquity of over- or under-dispersion in count data, we propose a Bayesian hierarchical modeling approach based on the renewal theory that relates nonexponential waiting times between events and the distribution of the counts, relaxing the assumption of equi-dispersion at the cost of an additional parameter. Particularly, we extend the methodology for the analysis of spatial count data based on the gamma distribution assumption for waiting times. The model can be formulated as a latent Gaussian model, and therefore, we can carry out the fast computation by using the integrated nested Laplace approximation method. The analysis of the groundwater dataset and a simulation study show a significant improvement over both Poisson and negative binomial models.
\end{abstract}

\begin{keywords}
Gamma-count distribution; Generalized Poisson likelihood; Integrated nested Laplace approximation; Renewal theory; Spatial count data; Stochastic partial differential equations.
\end{keywords}

\section{Introduction}
\subsection{Background}
All water collected beneath the surface of the earth is known as groundwater (Bear, 1979). Groundwater includes about $97$ percent of all the available freshwater on earth, excluding glaciers and ice caps (Hornberger et al. 2014).
In many countries, groundwater is the main source of drinking water. Especially, Iran suffers from the water shortage mainly due to insufficient rainfall and its desert environment. In such a dry and semi-dry country,  groundwater is often the only source of water available for drinking, agricultural, and industrial purposes. Groundwater quality could be affected by pollution arising from saltwater intrusion from desert or sea, use of chemical fertilizers, and the disposal of municipal sewage. 
Uncontrolled use of chemical fertilizers for agricultural purposes and infiltration of municipal wastewaters into groundwater due to the lack of sewage systems may have further reduced its quality and made it potentially hazardous to drink or use for the irrigation of crops (Vesali Naseh et al., 2018). Hence, it is essential to monitor the quality of groundwater continuously, to predict the potential health effects on the people originating from drinking and irrigating with this groundwater.

\begin{figure}[!htbp]
\begin{center}
\begin{tabular}{cc}
\includegraphics[width=5.1cm,height=5.1cm,keepaspectratio]{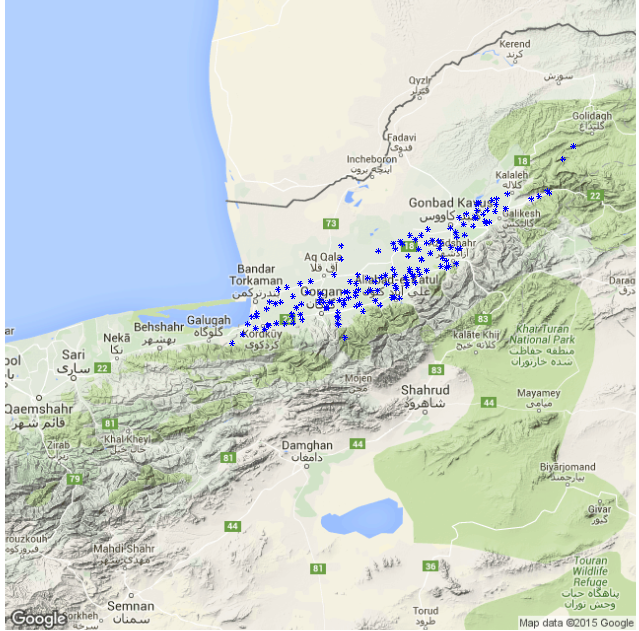} & \includegraphics[width=6.5cm,height=6.5cm,keepaspectratio]{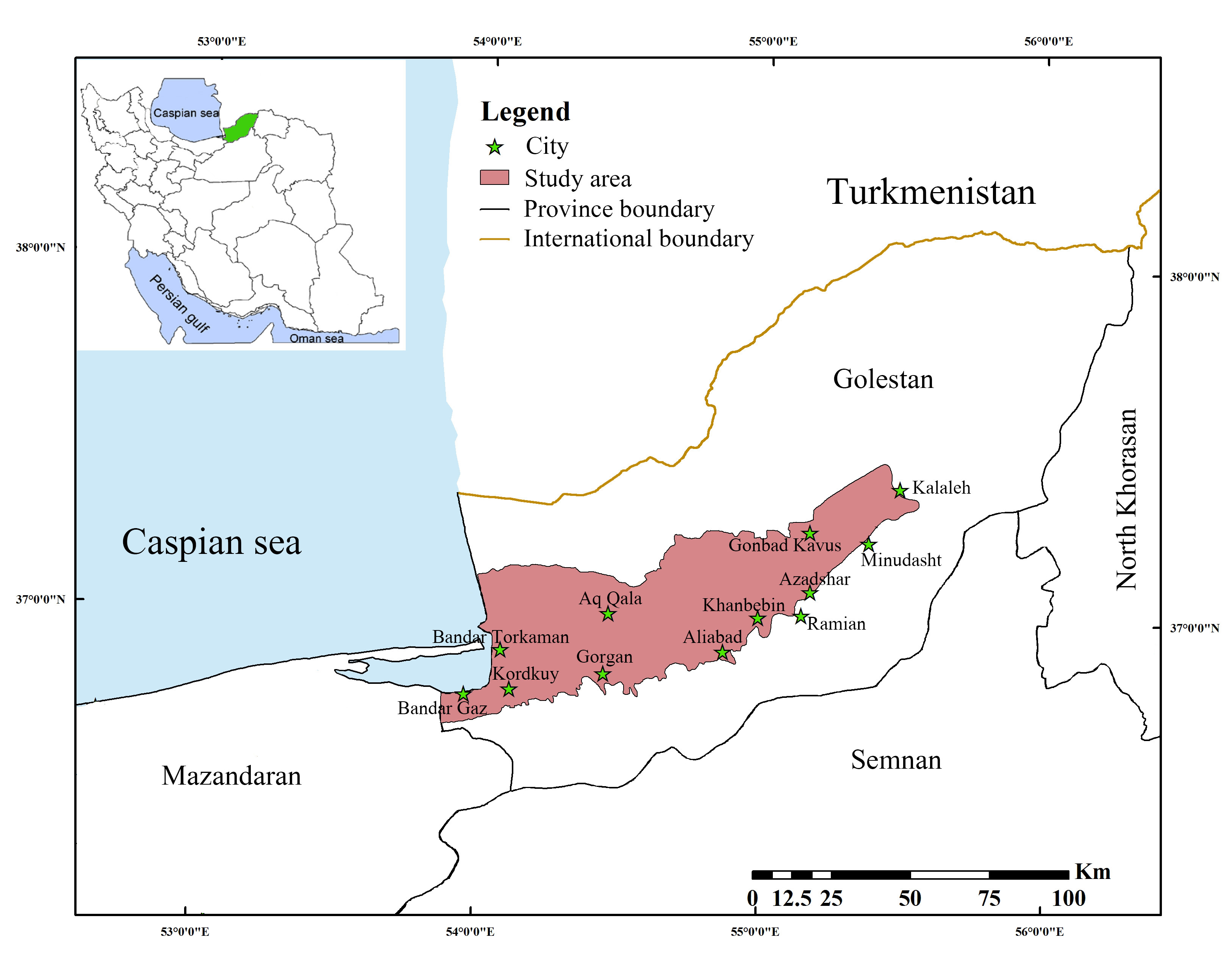}\\
\end{tabular}
\caption{\label{fig1} Locations of groundwater stations (blue stars) in Golestan province (left panel), a plot of the study area in Golestan Province (right panel)}
\end{center}
\end{figure}

The problem discussed in the present paper is motivated by analyzing geo-referenced measurements of groundwater collected from 150 springs, deep and semi-deep wells in a part of Golestan province, Iran (see Figure \ref{fig1}). Golestan is located in the north-east of Iran, the nearby Caspian Sea. Three different climates exist in the region: plain moderate, mountainous, and semi-arid. This province could also be considered as an agricultural and industrial region in Iran. 

For assessing the quality of groundwater, we use Electrical Conductivity (EC) that is measured by passing an electric current between two metal plates (electrodes) in the water sample and measuring how readily current flows (i.e., conducted) between the plates. This variable is measured in micromhos per centimeter (µmhos/cm) or microsiemens per centimeter (µs/cm). EC in water is affected by the presence of inorganic dissolved solids such as chloride, sulfate, sodium, and magnesium. Indeed, the more solids in the water, the stronger the current flow and the higher the EC. According to Lewis (2003), EC for groundwater stations supporting acceptable quality for a drink has a range below 800 µmhos/cm. Considering this threshold value for EC, our target response variable to monitor the quality of groundwater, given the presence of inorganic dissolved solids, is the number of counts that the quality of water is good for a drink from November 2003 to November 2013 that is reported by Water and Wastewater Company of Golestan Province. 

Due to the nature of groundwater quality measurements, the spatial variability of the response variable can be substantial. As we noted, values for EC vary for different climates, different inorganic dissolved solids properties, and various human behaviors, and hence typical values are region dependent. The modeling of the spatial variation of groundwater quality is vital for some reasons: 
\begin{enumerate}
\item[1)]
to display the intensity of acceptable quality of drinking water geographically for public advice and descriptive purposes.
\item[2)]
to inform to policy-makers for developing long-term water management strategies.
\item[3)]
to detect the most crucial dissolved solids in water that affect its quality. 
\end{enumerate}

\subsection{Modeling dispersed count data}
Spatial count data are observations from random variables with non-negative integer values for which each observation is spatially indexed, representing the number of times an event occurs. Such data may occur either from direct observation of quantities with integer values or as the result of the aggregation of observed quantities which denote the presence or absence of some characteristic; the latter is the case for our groundwater data. Modeling spatial count responses has many applications in numerous scientific disciplines such as environmetrics, ecology, geography, meteorology, biostatistics, and demography to name a few. A prevalent model for count data is the Poisson regression model as a member of generalized linear models (GLMs) family (Nelder and Wedderburn, 1972). However, in many situations, this model is not adequate due to the assumption of the equality of the conditional variance and mean which is too restrictive for (over or under) dispersed data. In particular, the conditional variance is higher than the conditional mean, that is named over-dispersion and vice versa under-dispersion. In either case, the Poisson regression model could result in biased inferences  (Winkelmann and Zimmermann, 1994).

For the groundwater data, we employed standard testing of the presence of over-dispersion in our count data. To this end, we considered the proposed classical test of Dean and Lawless (1989) for detecting over-dispersion in analyzing count data. They developed their proposed test as a score test against arbitrarily mixed Poisson alternatives. 
Let $\varsigma_1, \ldots, \varsigma_n$ be continuous positive-valued i.i.d. random variables such that, 
$Y_i |\bx_i, \varsigma_i \sim {\rm Poisson}(\varsigma_i \mu_i)$, where $\mu_i=\exp(\bx'_i\bbeta)$. Dean and Lawless (1989) assumed that the $\varsigma_i$'s have finite first and second moments, and without loss of generality they considered ${\rm E}(\varsigma)=1$ and ${\rm Var}(\varsigma)=\tau$. Then 
\[{\rm Var}(Y_i |\bx_i)=\mu_i+\tau \mu_i^2.\]
Therefore, a test for the Poisson model against a model with over-dispersion can be considered by testing $H_0: \tau=0$ versus $H_1: \tau >0$.
The proposed test statistic of Dean and Lawless (1989) is given by
\begin{eqnarray}\label{overstat}
T=\frac{\sum_{i=1}^{n}\left(\left(y_i -\hat{\mu}_i\right)^2-y_i\right)}{\sqrt{2\sum_{i=1}^{n}\hat{\mu}_i ^2}}
\end{eqnarray}
where $\hat{\mu}_i=\exp(\bx'_i\hat{\bbeta})$, with $\hat{\bbeta}$ the ML estimate of $\bbeta$ under the Poisson model.
Given the observed groundwater data, we computed the test statistic $T$ in \eqref{overstat} with observed value $-5.4466$. Based on the asymptotic normal distribution of $T$, the null hypothesis $H_0$ for these data is rejected at the level $0.05$; hence, we can conclude that the count data are overdispersed.

During the years, many approaches have been developed for dealing with over-dispersed count data such as adopting a generalized linear mixed model (GLMM) (Breslow and Clayton, 1993) by considering a random effect in the model. A usual proposal is a Poisson model with gamma distributed random effects leading to a negative binomial (NB) model. However, the negative binomial model is not a good alternative for under-dispersed data (Cameron and Trivedi, 2013). Two alternative classes of models for accounting unobserved heterogeneity are finite mixture models (Pearson, 1894) and hurdle models (see, for example, Cameron and Trivedi, 2013). The latter treats the modeling process for zeros differently from that for the non-zero counts. The hurdle model can be used for modeling both over-dispersion and under-dispersion. Other approaches include weighting the Poisson distribution (Ridout and Besbeas, 2004), the COM-Poisson distribution (Sellers and Shmueli, 2010, Lord et al.,  2010; Lord et al., 2008) and generalized Poisson inverse Gaussian family (Zhu and Joe, 2009) to name a few. Indeed, less attention has been paid for analysis of under-dispersed data in the literature (Zeviani et al., 2014; Ridout and Besbeas, 2004). Cameron and Triveidi (2013) is an excellent monograph for different count data models. 

In this paper, we consider the analysis of spatial over- or under-dispersed counts by using renewal theory (Cox, 1962) that relates nonexponential durations (waiting times) between events and the distribution of the counts. This methodology was first proposed by Winkelmann (1995). He connected the models for counts and models for durations relaxing the assumption of equi-dispersion at the cost of an extra parameter. Winkelmann replaced independently and identically exponentially distributed waiting times (which would lead to the Poisson distribution for counts) by a less restrictive nonnegative distribution with nonconstant hazard function. If the hazard function is a decreasing (increasing) function of time, the distribution reveals a negative (positive) duration dependence. It is shown that negative duration dependence causes over-dispersion and positive duration dependence under-dispersion (Winkelmann, 1995). Different researchers have proposed some models considering this methodology. The models of this type include gamma-count model (Winkelmann, 1995; Toft  \textit{et al.}, 2006), Weibull-count model  (McShane \textit{et al.},  2008), and lognormal-count model (Gonzales-Barron and Butler, 2011). Following a point from Winkelmann (1995), we extend the methodology for spatial count data analysis based on the gamma-count model:
\begin{quote}
Both Weibull and gamma nest the exponential distribution, and both allow for a (monotone) nonconstant hazard-that is, duration dependence. Although the Weibull distribution is preferred in duration analysis for its closed-form hazard function, the gamma distribution is preferred here for its reproductive property: Sums of independent gamma distributions are again gamma distributed.
\end{quote}

Although generalized coun data models based on renewal theory for analyzing count data could make a powerful technique, their use and applications have been less considered. A recent investigation is found in Zeviani \textit{et al.} (2014) who adopted the gamma-count model in the analysis of experimental under-dispersed data. They proposed a gamma-count regression model for implementing the maximum likelihood inference for uncorrelated counts. 

\subsection{A Bayesian framework}
Computational difficulties rendered likelihood-based inferences for a spatial gamma-count model cumbersome, that means computing the likelihood function of the model needed for such inferences requires calculating an intractable, high dimensional integral. Due to the advances in computation, especially Markov chain Monte Carlo (MCMC) sampling algorithms, we consider a Bayesian paradigm as the conventional approach for inference. It is well known, however, that MCMC algorithms come with a wide range of problems regarding convergence and computational time when applied to models like our proposed model. To overcome the difficulties associated with MCMC methods, Rue \textit{et al.} (2009) introduced a very fast, non-sampling based, approximate Bayesian methodology named integrated nested Laplace approximations (INLA). This method combines Laplace approximations and numerical integration in a very efficient manner for a particular class of models, the so-called latent Gaussian models. INLA substitutes MCMC simulations with accurate, deterministic approximations to posterior marginal distributions. The R package INLA (\texttt{R-INLA}) is free for download from %\url{http://www.r-inla.org/} 
\textit{http://www.r-inla.org/}
and fairly easy to use. Many examples of applications in several fields have appeared in the recent literature (S{\o}rbye \textit{et al.}, 2019; Bakka  \textit{et al.}, 2018; Rue \textit{et al.}, 2017; Muff \textit{et al.}, 2015; Blangiardo \textit{et al.}, 2013; Martins \textit{et al.}, 2013; Schr\"{o}dle and Held, 2011; Paul \textit{et al.}, 2010).

The INLA approach covers the family of Gaussian Markov random field (GMRF) models. A GMRF is a discretely indexed Gaussian random field that results in a sparse precision matrix for the model. This sparsity induces excellent computational properties which allow fast and accurate Bayesian inferences for latent Gaussian models (Rue and Held, 2005). However, for analysis of spatial point-referenced (geostatistical) data, the conventional model is a continuously indexed Gaussian random field (GRF). A GRF suffers computational issues due to the general cost of $\mathcal{O}(n^3)$ to factorize dense $n\times n$ (covariance) matrices. To solve this problem, we use a GRF approximation based on the theory of stochastic partial differential equations (SPDEs) that was introduced by Lindgren \textit{et al.} (2011). In the  SPDE approach, an explicit link between GRFs and GMRFs is provided by using a basis function representation with piecewise linear basis functions, and Gaussian weights with Markov dependences determined by a general triangulation of the domain. Along with the SPDE approach, the Bayesian inference in our application for groundwater quality assessment is carried out by the INLA method.

The plan of the rest of this paper is as follows. In Section \ref{Sec2}, we briefly explain the gamma-count regression as a flexible Poisson model. The essential methodology for Bayesian spatial gamma-count regression analysis for point-referenced data is developed in Section \ref{Sec3}. The performance of the proposed approach is examined in a simulation study in Section \ref{Sec4}. Section \ref{Sec5} applies the methodology to the groundwater quality dataset. We conclude the work in Section \ref{Sec6}.

\section{A Flexible Poisson Regression Model}\label{Sec2}
In this section, we briefly review the essential properties of the gamma-count (GC) distribution and its underlying classical regression model for uncorrelated data. 

As Winkelmann (2008) has noticed, the count and the duration view are just two different representations of the same underlying stochastic process. From a statistical viewpoint, the distributions of cumulative waiting times uniquely determine the distribution of counts and vice versa. This relationship can be employed to derive new count data distributions (Winkelmann, 1995; McShane \textit{et al.}, 2008, Gonzales-Barron and Butler, 2011; Ong \textit{et al.}, 2015). For example, the Poisson distribution corresponds to exponential interarrival times between events. The GC distribution has been proposed based on gamma distributed interarrival times by Winkelmann (1995).

Let $\tau_k$ be the waiting time between the $(k-1)$th and $k$th events. Therefore, the arrival time of the $n$th event is given by
\[\vartheta_n=\sum_{k=1}^n\tau_k,~~~n=1,2,\ldots .\]
Let $Y_T$ denote the total number of events that have occurred between $0$ and $T$. Hence, $\{Y_T, ~ T>0\}$ is a counting process and for a fixed $T$, $Y_T$ is a count variable. The stochastic properties of the counting process (and consequently of the count variable) are entirely determined once we know the joint distribution function of the waiting times, $\{\tau_k,~k\geq 1\}$. In particular, $Y_T<n$ if and only if $\vartheta_n>T$. Therefore,
\[{\rm P}(Y_t<n)={\rm P}(\vartheta_n>T)=1-F_n(T)\]
where $F_n(T)$ is the distribution function of $\vartheta_n$. Conversely, 
\[{\rm P}(Y_T=n)=F_n(T)-F_{n+1}(T).\]
Generally, $F_n(T)$ is a complicated convolution of the underlying densities of $\tau_k$'s, which makes it analytically intractable. However, by using the theory of renewal processes (Cox, 1962), a significant simplification arises if $\tau_k$'s are identically and independently distributed with a common distribution.

Here, we assume that $\{\tau_k,~k\geq 1\}$ is a sequence of independently and identically gamma distributed variables, $Gamma(\alpha, \gamma)$, with mean ${\rm E}(\tau)=\alpha/\gamma$ and variance ${\rm Var}(\tau)=\alpha/\gamma^2$.
It can be shown that if $Y_T$ denotes the number of events within $(0,T)$ interval, it is a GC distributed variable with parameters $\alpha$ and $\gamma$, denoted by $Y_T\sim GC(\alpha, \gamma)$. The probability mass function of $Y_T$ is given by 
\begin{eqnarray}\label{f2}
{\rm P}(Y_T=y)=G(y\alpha,\gamma T)-G((y+1)\alpha,\gamma T), ~~~~~ y=0,1,2,\ldots,
\end{eqnarray}
where
\[G(n\alpha,\gamma T)=\frac{1}{\Gamma(n\alpha)}\int_0^{\gamma T}v^{n\alpha-1}e^{-v}dv,~~~~~n=1,2,\ldots\]
and $G(0,\gamma T)=1$. For non-integer $\alpha$, no closed form expression is available for $G(y\alpha,\gamma T)$ and thus for ${\rm P}(Y_T=y)$.
However, for $\alpha$ taking integer values, \eqref{f2} is changed to
\begin{eqnarray}\label{f2.1}
{\rm P}(Y_T=y)={\rm e}^{-\gamma T}\sum_{j=0}^{\alpha-1}\frac{(\gamma T)^{\alpha y+j}}{(\alpha y+j)!}, ~~~~~ y=0,1,2,\ldots.
\end{eqnarray}
For $\alpha=1$, the distribution of $\tau$ reduces to the exponential, and \eqref{f2.1} simplifies to the Poisson distribution with the parameter $\gamma T$. More importantly, for positive duration dependence, $\alpha>1$ and the GC distribution is under-dispersed; for negative duration dependence, $0<\alpha<1$ and the GC distribution is over-dispersed. We refer readers to Winkelmann (1995) for more details about the definition and properties of the GC distribution.

For developing a GC regression model, we can relax the assumption of a homogeneous population by formulating a conditional model in which the mean of the count variable depends on a vector of covariates, $\bx=(1, x_1,\ldots,x_{p-1})^\prime$. The mean of GC distribution is
\begin{eqnarray}\label{f3}
{\rm E}(Y_T)=\sum_{k=1}^{\infty}G(k\alpha ,\gamma T),
\end{eqnarray}
that has no closed form. 
Assuming that the length of the time interval is the same for all observations, we can set $T$ to unity, without loss of generality. This results in the following regression model (Zeviani et al., 2014):
\begin{eqnarray}\label{f4}
{\rm E}(\tau_i|\bx_i)=\frac{\alpha}{\gamma_i}=\exp\{-\bx_i^{'}\bbeta\},
\end{eqnarray}
where $\bbeta$ is the $p\times 1$ vector of regression coefficients with the first element as the intercept.
We should notice that the regression model is defined on the waiting times $\tau_i$ instead of $Y_i$. Its origin is for failure to establish the equality ${\rm E}(Y_i|\bx_i)=\big({\rm E}(\tau_i|\bx_i)\big)^{-1}$ unless for $\alpha=1$. Indeed, given the inverse relationship between gaps and the number of occurrences, the minus sign behind $\bbeta$ is due to the reverse effect of covariates on waiting times instead of counts; the longer the expectation of time interval, the fewer the number of occurrences. Therefore, the GC regression model is developed from inherent parametric assumptions that nest the Poisson regression model by a singular parametric constraint. Succinctly, this model is a flexible kind of generalized Poisson model, which has some merits: 
\begin{enumerate}
\item[1)] It maintains a count data model with an additional parameter, which leads to higher flexibility than the Poisson model.  
\item[2)] It can convert to the Poisson model when the value of the shape parameter of waiting times is equal to 1. 
\item[3)] It explains over-dispersion and under-dispersion in terms of an underlying sequence of waiting times.
\end{enumerate}

From \eqref{f4}, one can write 
\begin{eqnarray*}
\gamma_i =\alpha\exp(\bx_i^{'}\bbeta).
\end{eqnarray*}
Therefore, given a sample of independent observations $\{(y_i,\bx_i),i=1,\ldots, n\}$, the GC regression model can be written as 
\begin{eqnarray*}
Y_i|\bx_i ;\alpha,\bbeta\sim GC(\alpha,\alpha\exp(\bx_i^{'}\bbeta)),~~~~i=1,\ldots ,n.
\end{eqnarray*}
The probability mass function of the response variable and its corresponding likelihood function are given by
\begin{eqnarray*}
f(y_i|\bx_i;\alpha,\bbeta)=G(y_i\alpha,\alpha\exp(\bx_i^{'}\bbeta))-G((y_i+1)\alpha,\alpha\exp(\bx_i^{'}\bbeta)),
\end{eqnarray*}
and
\begin{eqnarray*}
{\rm L}(\alpha,\bbeta|\y)&=&\prod_{i=1}^{n} \{G(\alpha y_{i},\alpha\exp(\bx_i^{'}\bbeta))-G(\alpha y_{i}+\alpha ,\alpha\exp(\bx_i^{'}\bbeta))\},
\end{eqnarray*}
respectively, where 
%$\bx=[\bx_1 , \ldots , \bx_n]$ is the matrix design and
 $\y=(y_1,\ldots,y_n)^\prime$ is the vector of observed counts.
There are no explicit forms for the maximum likelihood estimators of the parameters, due to the complexity of the likelihood function. Therefore, numerical optimization is needed for estimating the model parameters (Winkelmann, 1995; Zeviani \textit{et al.}, 2014).

\section{Spatial GC Regression Model}\label{Sec3}
In this section, we extend the hierarchical spatial modeling of geostatistical count data by considering a GC regression approach. This modeling approach allows the incorporation of spatially random effects to capture unobserved spatial heterogeneity or spatial correlation that cannot be explained by the available covariates. 

Let $D\subseteq \Re^d$, $d\geq 2$, be the region of interest, e.g. the whole area displayed in Figure \ref{fig1}, and for the $i$th observation, $i=1,\ldots,n$, let $s_i$ be a particular location within $D$. For spatial data, the random effect is
typically considered to be a spatial process which is modeled by a zero-mean Gaussian random field (GRF). Here, we assume $\phi_i=\phi(s_i)$ is the $i$th realization of the latent GRF $\phi(s)$ with an isotropic Mat\'{e}rn spatial covariance structure (Cressie, 1993; Stein, 1999) defined as below
\begin{eqnarray}\label{f9}
{\rm cov}(h)=\frac{\sigma^2}{2^{\nu-1}\Gamma(\nu)}(\kappa\|h\|)^{\nu}K_{\nu}(\kappa\|h\|),
\end{eqnarray}
where $\|h\|$ denotes the Euclidean distance between any two locations $s,s^\prime\in\Re^d$, $h=s-s^\prime$, $\Gamma(\cdot)$ is the gamma function, and $K_{\nu}(\cdot)$ is the modified Bessel function of the second kind of order $\nu$. To apply the SPDE approach, we need to consider a Mat\'{e}rn family for covariance structure. For the Mat\'{e}rn covariance function, $\sigma^2$ is the marginal variance, and $\nu$ measures the degree of smoothness which is usually fixed. In the INLA-SPDE methodology, for $d=2$, the smoothness is fixed at $\nu=1$. 
Further, $\kappa> 0$ is the scaling parameter with an empirical range $r=\sqrt{8\nu}/\kappa$ where the spatial correlation is close to 0.1 for all $\nu$ (Lindgren et al., 2011). For a fixed smoothing parameter $\nu$, the larger the value of $r$, the stronger the spatial correlation. 

Now, we introduce the spatial GC regression model. According to the INLA methodology, we consider the class of latent Gaussian models (Rue et al., 2009). These models construct a subclass of structured additive regression models (Fahrmeir and Tutz, 2001) in which the mean of the response variable, $\mu={\rm E}(Y)$, is linked to a structured additive predictor
\begin{eqnarray}\label{f6.1}
\eta_i= g(\mu_i)=\sum_{k=0}^{p-1}\beta_kx_{ki}+\sum_{j=1}^{n_f}f_j(v_{ji})+\epsilon_i
\end{eqnarray}
where $x_{0i}=1$, and all random components are assigned Gaussian priors. In \eqref{f6.1}, $\bbeta$ represent the linear effect of covariates $\bx$, the $\{f_j(\cdot)\}$ are unknown functions of the covariates $\bv=(v_1,\ldots,v_{n_f})^\prime$, and the $\epsilon_i$s are unstructured random effects. Specifically, the functions $\{f_j(\cdot)\}$ can be employed to explain the smooth effects of continuous covariates, as well as temporal, spatial or other underlying dependency structures in the data. We will focus on discussing the simplest case in which the predictor $\eta$ is a combination of the linear effect part, $\bx^\prime\bbeta=\sum_{k=0}^{p-1}\beta_kx_{k}$, and one $f_j(\cdot)$ that is modeled using a GRF, to avoid technicalities which are not necessary to understand the ideas. We let $\phi(\cdot)$ follow the GRF model assumed for a specific function $f_j(\cdot)$.

Given the latent random variables $\bphi=(\phi_1,\ldots,\phi_n)^\prime$, we suppose that the observations $\{y_i\}$ are conditionally independent and follow a GC distribution. Then, the spatial GC model is given by
\begin{eqnarray*}
{\rm E}(\tau_i |\bx_i,\phi_i)=\frac{\alpha}{\gamma_i}&=&\exp(-\eta_i) \\
&=&\exp(-\bx_i^{'}\bbeta-\phi_i),~~i=1,\ldots ,n.
\end{eqnarray*}
Hence,
\begin{eqnarray}
\label{f6}
Y_i|\bx_i,\phi_i,\alpha,\bbeta\sim GC\big(\alpha,\alpha\exp(\eta_i)\big),~~i=1,\ldots ,n.
\end{eqnarray}

As discussed in the introduction, a GRF experiences computational issues produced by factorizing dense covariance matrices. To overcome the computational burden, Lindgren et al. (2011) introduced an exciting approach that carries out the practical computations using a GMRF as the approximate solution to an SPDE. Certainly, the GRF $\phi(s)$ with the Mat\'{e}rn covariance function \eqref{f9} is a stationary solution to 
\begin{eqnarray}\label{f10}
\left(\kappa^2-\Delta\right)^{\frac{\zeta}{2}}\left(\tau \phi\left(s\right)\right)=\mathcal{W}(s),~~~~~~~~s\in D\subset\Re^d
\end{eqnarray}
where $\Delta$ is the Laplace operator, and $\tau$ controls the variance. The process $\mathcal{W}(s)$ is a spatial Gaussian white noise with unit variance. The link to the Mat\'{e}rn smoothness $\nu$ and variance $\sigma^2$ is $\zeta=\nu + d/2$ and
\[\sigma^2=\frac{\Gamma(\nu)}{\Gamma(\zeta)(4\pi)^{d/2}\kappa^{2\nu}\tau^2}.\]
An integer $\zeta$ must be chosen to obtain a GMRF. By considering $\nu=1$, for $d=2$, we get $\zeta=2$ and $\sigma^2=1/(4\pi\kappa^2\tau^2)$. Obviously, two parameters $\kappa$ and $\tau$ have a joint
influence on the marginal variance of the latent field $\phi(s)$.

An approximation to the solution of the SPDE in \eqref{f10} can be obtained by using a finite element method defined on a triangulation of the domain $D$:
\[\phi(s)=\sum_{k=1}^m\psi_k(s)w_k.\]
Here, $\{w_k\}$ are the random weights chosen so that the distribution of $\phi(s)$ approximates the distribution of the solution to the SPDE on the domain, and $\{\psi_k\}$ are basis functions. In this representation, $m$ is the number of vertices in the triangulation, and the basis functions are chosen to be piecewise linear on each triangle, i.e., $\psi_k$ is 1 at vertex $k$ and 0 elsewhere. These piecewise linear functions induce a Markov structure. The Markov property allows a sparse precision matrix so that efficient numerical algorithms can be applied for large spatial data (Rue and Held, 2005). The projection of the SPDE onto the basis representation is chosen by a finite element method. See Lindgren et al. (2011) for more details.

Lindgren et al. (2011) proved that for $\zeta=2$ the random latent variables $\bphi$ construct a GMRF with mean 0, and its precision matrix $Q=Q(\bvarsigma)$ is given by
\[Q=\tau^2(\kappa^4\tilde{C}+2\kappa^2G+G\tilde{C}^{-1}G),\]
with elements of $\tilde{C}$ (a diagonal matrix) and $G$ as follows
\begin{eqnarray*}
\tilde{C}_{ii}&=& \int_D \psi_k(s)ds,\\
G_{ij} &=& \int_D \nabla\psi_i(s)^\prime \nabla\psi_j(s) ds
\end{eqnarray*} 
respectively. Here, $\bvarsigma=(\tau,\kappa)^\prime$ is the vector of parameters of the Mat\'{e}rn covariance structure.
The marginal likelihood function of the model \eqref{f6} is
\begin{eqnarray}
\label{f7}
{\rm L}(\alpha,\bbeta, \bvarsigma |\y)&=&\int_{\Re^n}\prod_{i=1}^nf(y_i,\phi_i|\alpha,\bbeta,\bvarsigma)
d\bphi \cr
&=&\int_{\Re^n}\prod_{i=1}^nGC\big(y_i;\alpha,\alpha\exp(\eta_i)\big)\mathcal{N}_n(\bphi ;\zero ,Q^{-1}(\bvarsigma))d\bphi.
\end{eqnarray}
Calculating the marginal likelihood \eqref{f7} nearly always involves intractable integrals, which is the central impediment for implementing likelihood-based inferences. Also, the computational difficulty increases with the number of observations due to this fact that the dimension of the spatial random variable is equal to the number of observations.

\subsection{Bayesian analysis of spatial GC model}
For analysis of spatial GC regression model, in a Bayesian framework, it is necessary to choose some suitable prior distributions for parameters of the model, $\alpha$, $\bbeta$, $\tau$, and $\kappa$, that can reflect our prior beliefs about them. We suppose that the parameters are a priori independent, that is $\pi(\alpha, \bbeta, \tau,\kappa)=\pi(\alpha)\pi(\bbeta)\pi(\tau)\pi(\kappa)$. The vector of parameters $\bbeta$ is assumed to have independent zero-mean Gaussian priors with fixed large variances. This choice is based on the INLA methodology (Rue et al., 2009). As $\alpha$ is a positive scale parameter, we consider a gamma distribution as the prior distribution for it. We also consider normal priors for both $\log(\tau)$ and $\log(\kappa)$. These prior distributions are flexible enough to represent the prior belief via the appropriate choice of their hyperparameters. 

Instead of using the default parameterization according to $\tau$ and $\kappa$, the SPDE methodology prefers to control the parameters through the marginal standard deviation, $\sigma$, and the range, $r$. For details on this parameterization see Lindgren (2012). Krainski et al. (2018) proposed to use the joint Penalized Complexity (PC) prior for these parameters, as derived in Fuglstad et al. (2019). The PC approach is a new framework for constructing prior distributions introduced by Simpson et al. (2017). The PC priors have important properties including invariant to reparameterizations, natural connection to Jeffreys' priors, support Occam's razor, and to have excellent robustness features. 

Following Fuglstad et al. (2019), a prior for $\sigma$ is defined by specifying parameters $(\sigma_0,q_1)$ so that
\[{\rm P}(\sigma>\sigma_0)=q_1,~\sigma_0>0,~0<q_1<1.\]
Similarly, for the range parameter we have to choose $r_0$ and $q_2$ shuch that
\[{\rm P}(r<r_0)=q_2,~r_0>0,~0<q_2<1.\]
The defined PC priors are weakly informative and penalize complexity of the model by shrinking the range towards infinity and the marginal variance towards zero. It is done by hyperparameters that show
how strongly the user demands to shrink towards the base model.

By accepting these priors, the hierarchical Bayesian spatial GC model can be written as
\begin{eqnarray}\label{f11}
Y_i|\bx_i, \phi_i, \alpha , \bbeta &\sim & GC\left(\alpha , \alpha \exp(\eta_i)\right),~~~i=1,\ldots ,n \nonumber \\
\bphi|\tau, \kappa &\sim & \mathcal{N}_n(\zero,Q^{-1}) \nonumber\\
\alpha & \sim & {\rm Gamma}(a,b)\nonumber\\
\bbeta & \sim & \mathcal{N}_p(\zero,\Sigma_{\beta})\nonumber \\
\log(\tau) & \sim & \mathcal{N}(c,d)~~or~~\sigma\sim{\rm PC}(\sigma_0, q_1)\nonumber\\
\log(\kappa) & \sim & \mathcal{N}(e,f)~~or~~r\sim{\rm PC}(r_0,q_2) 
\end{eqnarray}
where $\eta_i=\bx_i^\prime\bbeta+\phi_i$, and $a, b, \Sigma_\beta, c, d, e$, and $f$ are known hyperparameters of the prior distributions. Some default values for them are provided in the INLA methodology. 
The joint posterior density is now as follows:
\begin{eqnarray*}
\pi(\alpha,\bbeta,\tau,\kappa,\bphi|\y)&\propto &{\rm L}(\alpha,\bbeta,\tau,\kappa|\y) \pi(\bphi|\tau,\kappa)\pi(\alpha) \pi(\bbeta)\pi(\tau)\pi(\kappa)\cr
&\propto &\prod_{i=1}^{n} \left\{G\left(\alpha y_{i},\alpha\exp\left(\eta_i\right)\right)-G\left(\alpha y_{i}+\alpha ,\alpha\exp\left(\eta_i\right)\right)\right\}\cr
&\times & |Q|^{\frac{1}{2}}\exp\left\{-\frac{1}{2}\bphi^{\prime}Q\bphi\right\}\cr
& \times &\alpha^{a-1}\exp\left(-b\alpha\right)\cr
&\times & \exp\left\{\bbeta^\prime\Sig^{-1}_\beta\bbeta\right\}\cr
&\times & \pi(\tau)\pi(\kappa).
\end{eqnarray*}

The conventional approach to inference for the model \eqref{f11} is based on MCMC sampling. It is well known, however, that MCMC methods have serious problems, regarding both convergence and computational time, when applied to such models (Rue et al., 2009). Particularly, the complexity of the proposed model for large spatial data could lead to several hours or even days of computing time to implement Bayesian inference via MCMC algorithms. To overcome this issue, Rue et al. (2009) introduced the INLA method that is a deterministic algorithm and provides accurate results in seconds or minutes. INLA combines Laplace approximations (Tierney and Kadane, 1986) and numerical integration in a very efficient manner to approximate posterior marginal distributions. Let $\btheta=(\alpha, \tau, \kappa)^\prime$ denote the hyperparameters of the model \eqref{f11}. Let also $\bpsi=(\bbeta,\bphi)^\prime$ denote the $s\times 1$ vector of latent variables where $s=p+n$. In practice, the primary interest lies in the marginal posterior distributions for each element of the latent variables vector
\[\pi(\psi_j|\y)=\int\pi(\psi_j,\btheta|\y)d\btheta=\int\pi(\psi_j|\btheta,\y)\pi(\btheta|\y)d\btheta,~~j=1,\ldots,s,\]
and for each element of the hyperparameter vector
\[\pi(\theta_k|\y)=\int\pi(\btheta|\y)d\btheta_{-k},~~k=1,2,3,\]
where $\btheta_{-k}$ is equal to $\btheta$ with eliminated $k$th element. The essential feature of INLA is to use this form to construct nested approximations
\begin{eqnarray*}
\tilde{\pi}(\psi_j|\y)&=&\int\tilde{\pi}(\psi_j|\btheta,\y)\tilde{\pi}(\btheta|\y)d\btheta, \\
\tilde{\pi}(\theta_k|\y)&=&\int\tilde{\pi}(\btheta|\y)d\btheta_{-k},
\end{eqnarray*}
where Laplace approximation is applied to carry out the integrations required for evaluation of $\tilde{\pi}(\psi_j|\btheta,\y)$.

A crucial success of INLA is its ability to compute model comparison criteria, such as deviance information criterion (DIC; Spiegelhalter et al., 2002) and Watanabe-Akaike information criterion (WAIC; Watanabe, 2010; Gelman et al. 2014), and various predictive measures, e.g., conditional predictive ordinate (CPO; Pettit, 1990) and probability integral transform (PIT; Dawid, 1984), to compare different possible models. Our proposed GC model has already implemented in the \texttt{R-INLA} package as a \texttt{family} argument with the name "\texttt{gammacount}".

\section{Simulation Study}\label{Sec4}
We illustrate the performance of the proposed GC regression in modeling geostatistical counts using the SPDE approach with simulated data. Due to the primary purpose of geostatistics modeling, i.e., spatial prediction, we mostly aim to assess the predictive performance of our proposed GC model. To this end, we judge the performance of the model based on the ability to recover the true parameter values, cross-validated predictive measures, and visual similarity between the predicted and true spatial effect surfaces. Then, we compare the results to the ones obtained from spatial Poisson and negative binomial (NB) regression models as the traditional counterparts.

Let $\phi(s),~s\in [0,1]\times [0,1]$, be a mean-zero GRF with the Mat\'{e}rn covariance function \eqref{f9}. We used the SPDE representation \eqref{f10} for $\phi(s)$ and generated the realization $\bphi=(\phi_1,\ldots,\phi_{200})^\prime$ of the field with parameters $\nu=1$, $\sigma^2=1$, and $r=0.2$. We modeled the linear predictor as
\[\eta_i=\vartheta_0 + \vartheta_1 x_i + \phi_i,~~i=1,\ldots,200,\]
in which the covariate $x$ was generated from a standard normal distribution. We set the real values for the regression parameters at $\vartheta_0=1$ and $\vartheta_1=-0.7$. Since the parameter $\alpha$ indicates the various dispersion conditions in the GC model, we considered a range of variations for this parameter in $\{0.1, 1, 1.5, 3\}$. It is useful to remind that $\alpha=1$, $\alpha>1$, and $\alpha<1$ are corresponding to equivalent dispersion, under-dispersion, and over-dispersion, respectively. Furthermore, a GC model with $\alpha=1$ corresponds to a Poisson model.

We computed several measures to evaluate the performance of the three alternative models. The efficiency of the estimators of parameters for proposed models is assessed by the root mean squared errors (RMSE), which is defined as
\begin{eqnarray*}
{\rm RMSE}(\hat{\delta}) =\sqrt{ \frac{1}{R}\sum_{r=1}^{R}(\hat{\delta}_r - \delta)^2}
\end{eqnarray*}
where $R$ expresses the number of replications, fixed here at $R=1000$. Also, we computed WAIC as a predictive cross-validated measure to compare the models. This criterion is recommended by Gelman et al. (2014) over the DIC, as a fast and computationally-convenient alternative with some better properties. A smaller value of WAIC implies better model characteristics.

To estimate the proposed models, we considered the prior distributions given in \eqref{f11} with $a=0.01$, $b=0.01$, $\Sigma_{\beta}=1000\times {\rm I}_3$, $c=\log(0.02)$, $d=10$, $e=\log(14)$, and $f=10$. Table \ref{Tab1} shows the RMSE values of the estimators of parameters for the suggested models under different amounts of dispersion. We observe that the GC model, in the estimation of both regression and covariance parameters, is uniformly more efficient than the other two alternatives when responses are under-dispersed or over-dispersed. It is, essentially, interesting when we recall that the NB model could take into account the potential overdispersion of the data. For each of the dispersion values, regression coefficients are estimated with acceptable accuracy for the GC model compared with using the Poisson and NB models. Reduced accuracy is seen for the NB model, especially when data are under-dispersed. Hence, the results show that the GC model (except for $\alpha=1$) has a better description of the real systematic trend of the spatial data than both Poisson and NB models.
The same results could be extracted for the parameters of the Mat\'{e}rn covariance function, where the GC model performs reasonably reliable and consistent in recovering actual values. For equi-dispersed data, the (correct) Poisson model slightly outperforms others.
\begin{table}[ppt]
%\begin{footnotesize}
\begin{center}
\caption{\label{Tab1} \footnotesize{The RMSE values of the estimators of parameters for the GC model along with their correspond values for the Poisson and negative binomial models$^\dag$}}
\begin{tabular}{cccccc}
\hline
& &\multicolumn{4}{c}{$\alpha$ (Dispersion parameter)} \\
\cline{3-6}   
Coefficient & Model & 0.1 & 1 & 1.5 & 3 \\
\hline 
& GC & \textbf{0.3576} & 0.2202 & \textbf{0.2399} & \textbf{0.2365}\\
$\vartheta_0$ & Poisson & 0.7122 & \textbf{0.2190} & 0.2495 & 0.2949 \\
& NB & 0.9598 & 20.1498 & 10.8487 & 2.0589 \\ \\
& GC & \textbf{0.1471} & 0.0620 & \textbf{0.0554} & \textbf{0.0694} \\
$\vartheta_1$ & Poisson & 0.2808 & \textbf{0.0584} & 0.0592 & 0.0808 \\
& NB & 0.2876 & 5.6965 & 5.9162 & 7.7255 \\ \\
& GC & \textbf{0.0772} & 0.0511 & \textbf{0.0553} & \textbf{0.0516} \\
$r$ & Poisson & 0.1365 & \textbf{0.0494} & 0.0615 & 0.0774 \\
& NB & 0.1155 & 0.0516 & 0.0631 & 0.0787 \\ \\
& GC & \textbf{0.3000} & \textbf{0.3169} & \textbf{0.3196} & \textbf{0.4785} \\
$\sigma$ & Poisson & 0.9351 & 0.7792 & 0.7757 & 0.7488 \\
& NB & 0.7545 & 0.7903 & 0.7838 & 0.7473 \\ \\
$\alpha$ & GC &  0.1146 & 0.1433 & 0.3288 & 0.9099\\
\hline
\end{tabular}
\end{center}
$^\dag$ \footnotesize{Bold values indicate the best selected one.}
%\end{footnotesize}
\end{table}

We also provided the preference rate (PR) of GC to Poisson and negative binomial models based on WAIC. PR is the proportion of selecting GC as the best model compared to the other two proposed models when the reference measure is WAIC. Table \ref{Tab2} summarises the PRs obtained under different models. As we expected, the result for PR(GC/Poisson) when $\alpha=1$ is in favor of the Poisson model. However, the results are in favor of the GC model when data are not equi-dispersed. Indeed, larger values of PR suggest that the Poisson or NB models do not fit the data as adequately as the GC model. Also, it shows that the GC model has better predictive nature in comparison to the competing models. Specifically, for under-dispersed data, the GC model strongly outperforms both Poisson and NB models.
\begin{table}[ppt]
\centering \caption{\label{Tab2} \footnotesize{Preference rates of the GC model to Poisson and NB models based on WAIC }}
\begin{tabular}{ccccc}
\hline
& \multicolumn{4}{c}{$\alpha$ (Dispersion parameter)} \\
\cline{2-5}
& 0.1 & 1 & 1.5 & 3\\
\hline
PR(GC/Poisson) & 0.542 & 0.386 & 0.892 & 0.970\\
PR(GC/NB) & 0.978 & 0.944 & 0.974 & 0.972\\
\hline
\end{tabular}
\end{table}
 
\begin{figure}[!htbp]
\begin{center}
\begin{tabular}{cccc}
\includegraphics[scale=0.15]{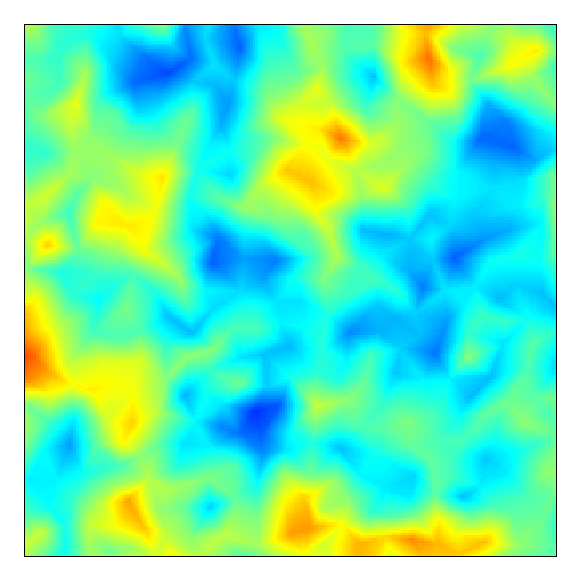} & \includegraphics[scale=0.15]{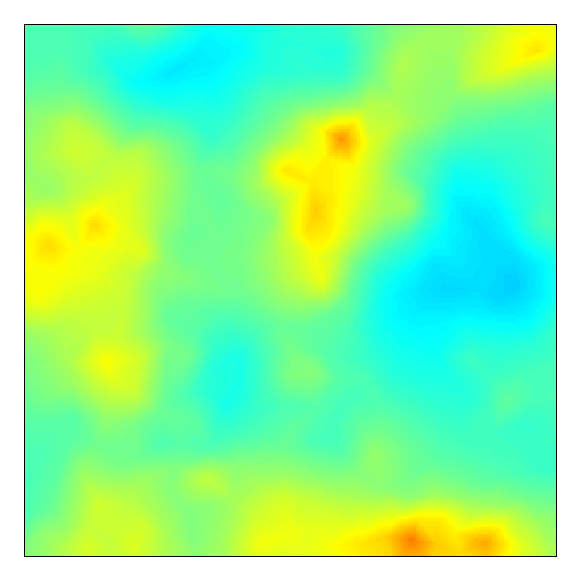}&\includegraphics[scale=0.15]{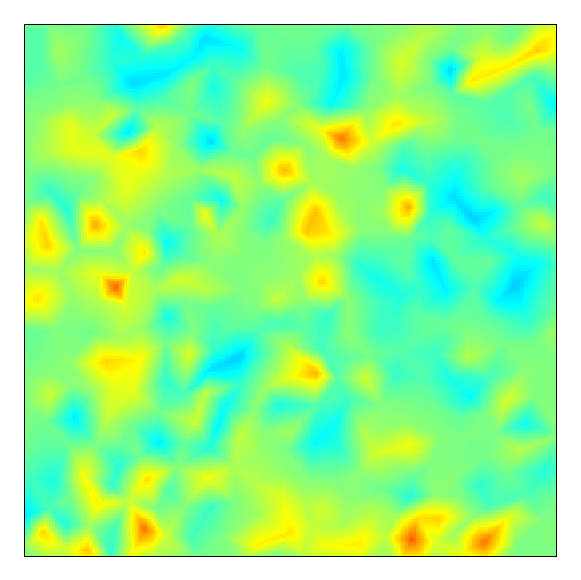} & \includegraphics[scale=0.15]{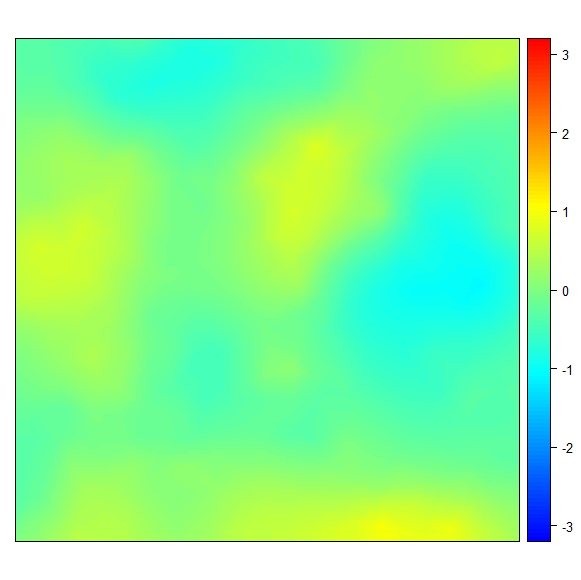}\\
\includegraphics[scale=0.15]{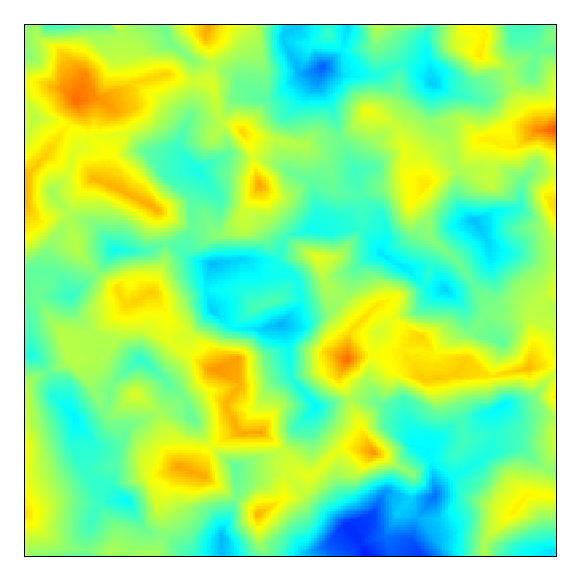} & \includegraphics[scale=0.15]{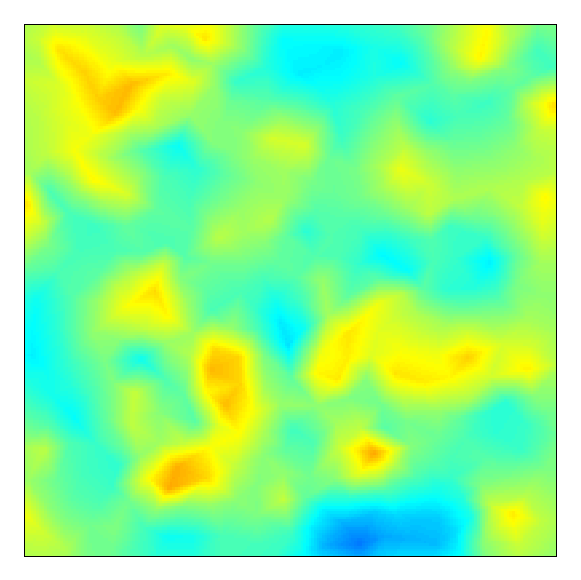}&
 \includegraphics[scale=0.15]{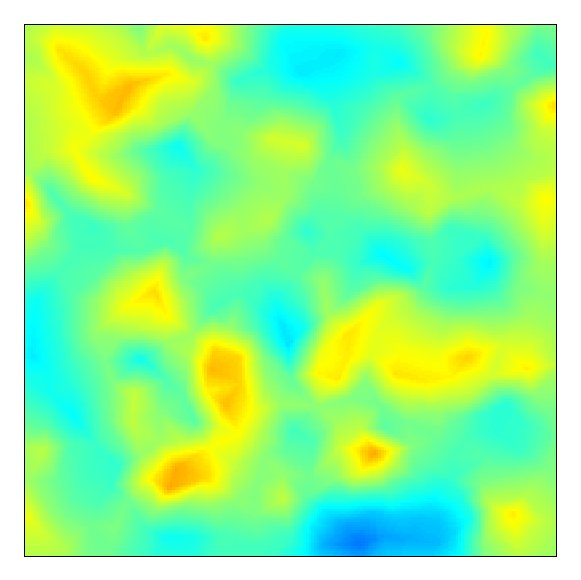} & \includegraphics[scale=0.15]{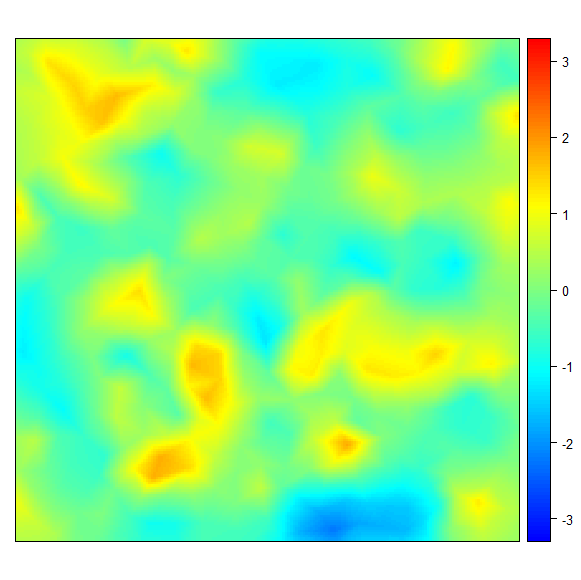}\\
 \includegraphics[scale=0.15]{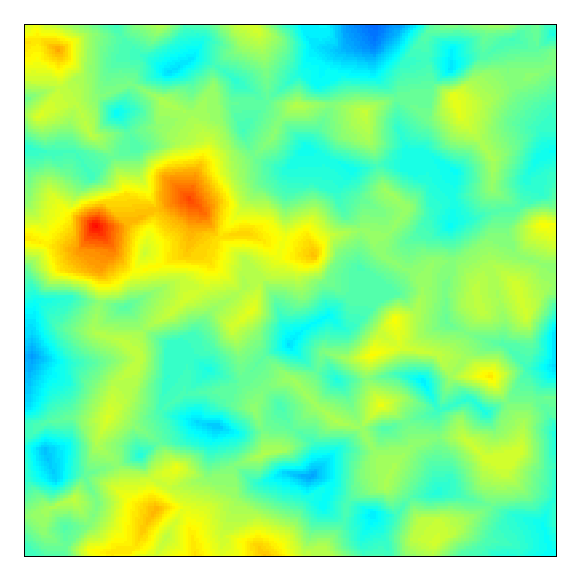} & \includegraphics[scale=0.15]{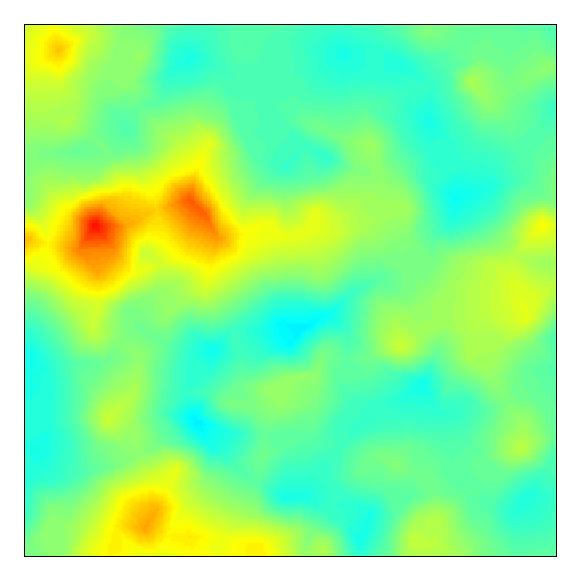}&
 \includegraphics[scale=0.15]{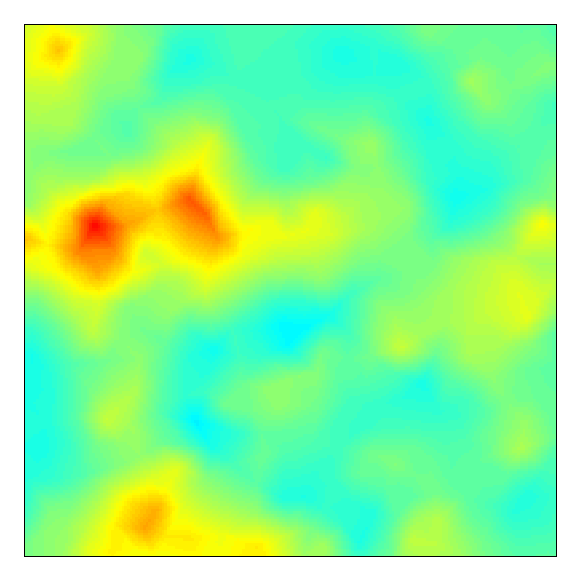} & \includegraphics[scale=0.15]{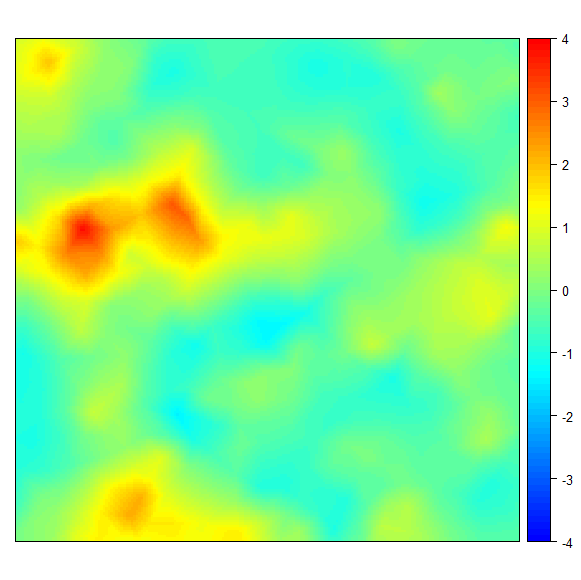} \\
  \includegraphics[scale=0.15]{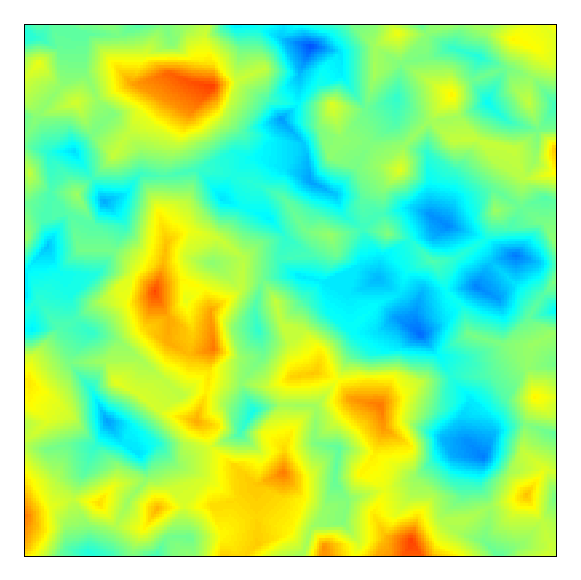} & \includegraphics[scale=0.15]{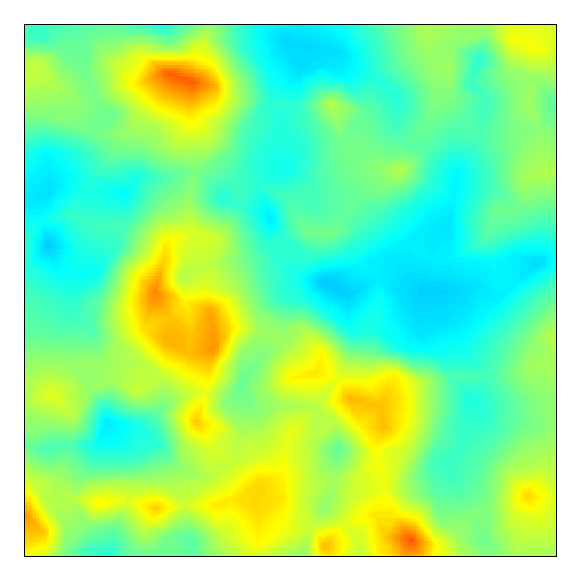}&
 \includegraphics[scale=0.15]{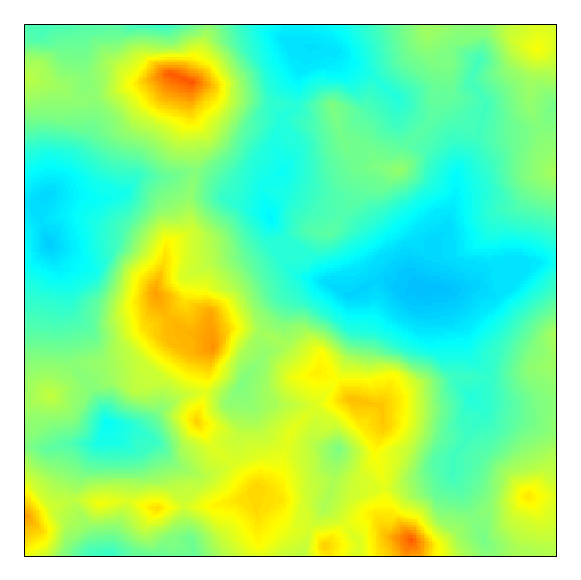} & \includegraphics[scale=0.15]{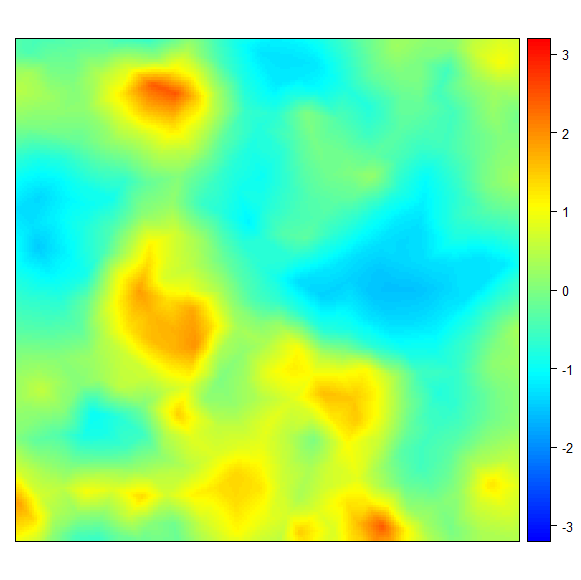}
\end{tabular}
\caption{\label{fig2} \footnotesize{A simulated Mat\'{e}rn field (first column), and the prediction (posterior mean) of the field based on GC model (second column), Poisson model (third column), and NB model (forth column). The results are obtained for $\alpha=0.1$ (first row), $\alpha=1$ (second row), $\alpha=1.5$ (third row), $\alpha=3$ (fourth row)}}
\end{center}
\end{figure}

We measured the prediction accuracy for the whole GRF with mean squared prediction error 
\[{\rm MSPE}=\sum_j^{n}(\hat{\phi}(s_i)-\phi(s_i))^2/n,\]
where $\phi(s_i)$ is the true value of one simulated GRF on location $s_i$ and $\hat{\phi}(s_i)$ is the prediction using corresponding posterior means. We computed the WAIC for each model, in this case, as well. Both criteria are useful for assessing prediction accuracy. Table \ref{Tab3} displays the results. The GC model outperforms the Poisson and NB models unless the data are equi-dispersed in which the Poisson model is superior according to the MSPE. Also, the interpolated surfaces of the simulated field, as well as the prediction of the field for three competing models, can be seen in Figure \ref{fig2} for different amounts of dispersion.  The GC model dominates the other two models in terms of recovering the true underlying spatial structure when data are under or over-dispersed. For equi-dispersed data, the performances of the three models are mostly equivalent.
\begin{table}[ppt]
%\begin{footnotesize}
\begin{center}
\caption{\label{Tab3} \footnotesize{MSPE and WAIC (${\rm MSPE|WAIC}$) using three competing models in SPDE example for different values of $\alpha^\dag$}}
\begin{tabular}{ccccc}
\hline   
 &\multicolumn{4}{c}{$\alpha$ (Dispersion parameter)} \\
\cline{2-5}
Model & 0.1 & 1 & 1.5 & 3\\ 
\hline
GC & {\bf 0.852}$|${\bf 1728.325} & 0.603$|${\bf 1278.023} & {\bf 0.631}$|${\bf 1137.329} & {\bf 0.422}$|${\bf 1053.406} \\ \\
Poisson & 1.041$|$1743.847 & {\bf 0.599}$|$1278.617 & 0.639$|$1174.026 & 0.429$|$1194.853 \\ \\
NB & 0.873$|$1731.386 & 0.601$|$1288.732 & 0.641$|$1183.766 & 0.428$|$1199.362 \\
\hline
\end{tabular}
\end{center}
$^\dag$ \footnotesize{Bold values identify the selected model.}
%\end{footnotesize}
\end{table}

\begin{figure}[!htbp]
\begin{center}
\begin{tabular}{cccc}
%$\alpha$=0.1 & $\alpha$=1 & $\alpha$=1.5 & $\alpha$=3 \\
\includegraphics[scale=0.15]{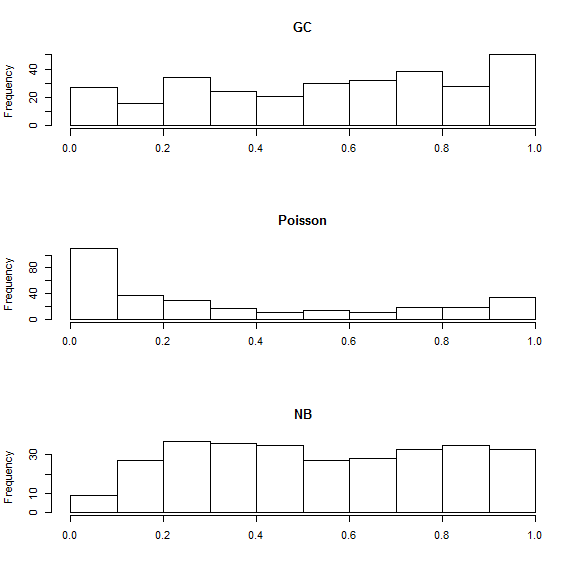} & \includegraphics[scale=0.15]{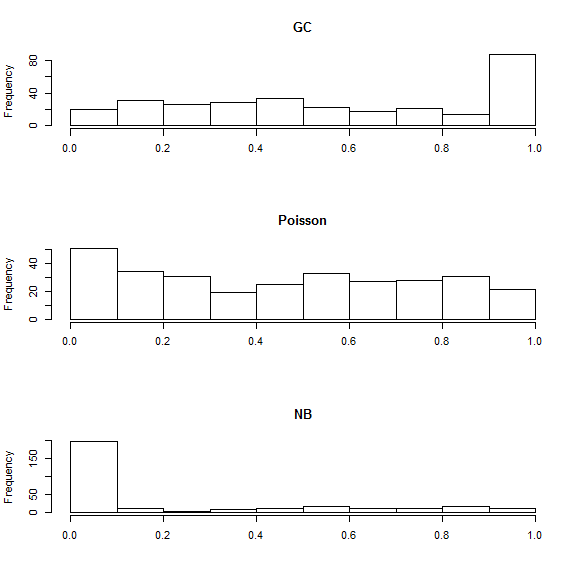}&\includegraphics[scale=0.15]{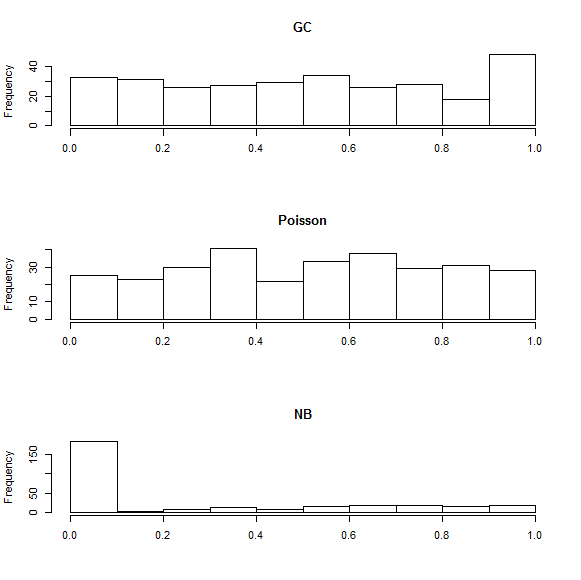} \includegraphics[scale=0.15]{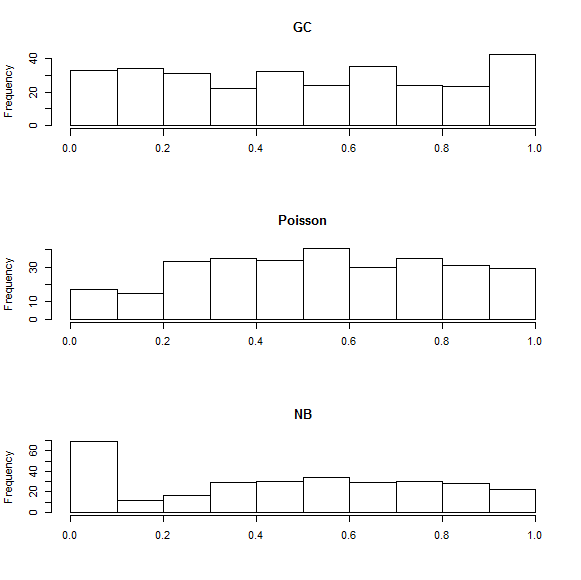}\\
\end{tabular}
\caption{\label{fig3} \footnotesize{Histograms of PIT values for the GC model (first row), the Poisson model (second row), and the NB model (third row). The results are obtained for $\alpha=0.1$ (first column), $\alpha=1$ (second column), $\alpha=1.5$ (third column), $\alpha=3$ (fourth column).}}
\end{center}
\end{figure}

Finally, to complete our evaluation, regarding prediction power, we computed the cross-validated PIT as well. The PIT is a leave-one-out cross-validation score that is given by
\[{\rm PIT}_i={\rm P}(Y_i<y_{i}^{obs}|\y_{-i})\]
where $\y_{-i}$ is the observations vector $\y$ with the $i$th component removed. For a well-calibrated model, the PIT values should be uniformly distributed. Histograms of the PIT values can, therefore, be used to assess the calibration of the model. As Figure \ref{fig3} indicates, the GC model is preferable than two other alternatives in various dispersion situation.


\section{Groundwater quality assessment}\label{Sec5}
In this section, we apply the proposed models to analyze the groundwater data for the 150 measurement stations in a north-east province of Iran, Golestan, shown in Figure \ref{fig1}.
As we noted in the Introduction, the number of counts that the quality of water is suitable for a drink (based on EC measures) was used as the response variable for the proposed models. The explanatory variables (covariates) in our study are given in Table \ref{Tab4}. 
\begin{table}[h!]
\begin{center}
\caption{\label{Tab4}\footnotesize{List of explanatory variables}}
\begin{tabular}{ccc}
\hline
variable & name & unit\\
\hline
Potassium & k & ${\rm mg~L^{-1}}$ \\
%\hline
Total Hardness & th & ${\rm mg~L^{-1}}$ \\
%\hline
Sodium & na & ${\rm mg~L^{-1}}$ \\
%\hline
Chloride & cl & ${\rm mg~L^{-1}}$ \\
%\hline
Bicarbonate & ${\rm hco}_3$ & ${\rm mg~L^{-1}}$ \\
%\hline
Total Dissolved Solids & tds & ${\rm mg~L^{-1}}$ \\
\hline
\end{tabular}
\end{center}
\end{table}
The observed measures for each covariate are the average of the yearly totals. The period from November 2003 to November 2013 was used to compute the annual average of the variables.

\begin{figure}[!htbp]
\begin{center}
\includegraphics[scale=0.6]{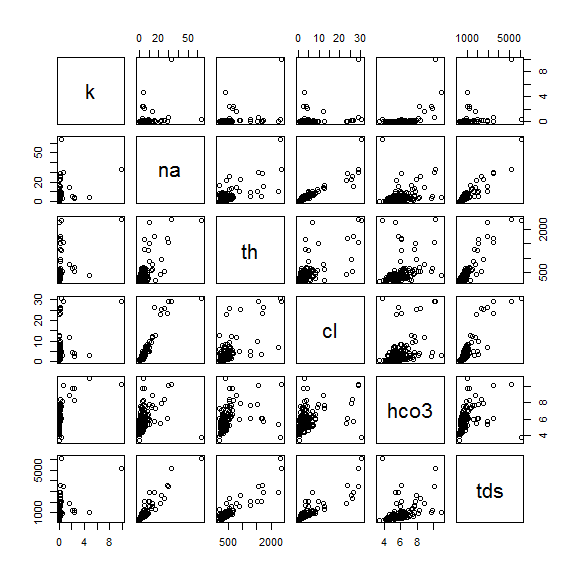} 
\caption{\label{fig4} \footnotesize{The matrix of scatter plots of covariates for groundwater data}}
\end{center}
\end{figure}

In Figure \ref{fig4}, the matrix of scatter plots of the covariates confirm that some of them are highly correlated; therefore adjusting for a set of covariates may indirectly adjust for other correlated ones as well, at least to some extent. This behavior could give some evidence of the best-selected model in terms of model selection criteria, as we will present in the following.

%\subsubsection*{Testing for overdispersion}

\subsection*{Fitting models}
\begin{figure}[!htbp]
\begin{center}
\includegraphics[scale=0.5]{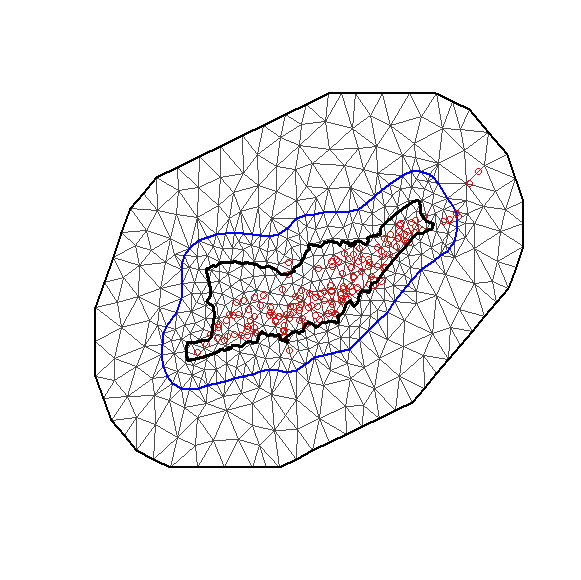} 
\caption{\label{fig5} \footnotesize{The Golestan province triangulation. The red dots denote the observation locations of the groundwater stations.}}
\end{center}
\end{figure}

The first step required to fit the geostatistical model for groundwater data, by the SPDE approach, is the triangulation of the considered spatial domain depicted in Figure \ref{fig1}. Figure \ref{fig5} displays a non-convex triangulation for the area study inside the Golestan province. To increase the approximation accuracy and avoid edge effect issues, we considered an expansion of the triangulation beyond the borders of the study region.

We performed multiple covariate analyses, by using the SPDE modeling approach, to select the best subset of covariates, adjusting for spatial random effects. We used the same hyperparameters as the simulation example for the prior distribution of the dispersion parameter of the GC model, $\alpha$, and regression parameters, $\bbeta$. 
We also used the default prior distributions for the Mat\'{e}rn field parameters of the model as proposed by Lindgren et al. (2011) and described in \eqref{f11}. From our results based on the GC model, shown in Table \ref{Tab5}, our analysis will be centered on the model with only {\it sodium} and {\it total hardness} as the covariates. This model is the best fitting model, according to all WAIC, DIC, and the cross-validated logarithmic score, Log.score, criteria. The Log.score criterion is computed by using CPO as follows:
\[{\rm Log.score} = -\sum_{i}\log({\rm CPO}_i)\]
where for each observation, CPO is given by
\[{\rm CPO}_i=\pi (y_{i}|\y_{-i})~~~~~~i=1,\ldots,n.\]
The CPO for the $i$th observation denotes the posterior probability of observing the value of $y_i$ when the model is fitted to all data without $i$th observation. A smaller value of the logarithmic score indicates a better prediction quality of the model.
The same results (not shown here) were obtained for both Poisson and NB models. The (posterior mean) estimates of the Mat\'{e}rn field parameters, for the selected model, are $\hat{\sigma}=0.74$ and $\hat{r}=49.18$. We used these estimates to set the hyperparameters in the PC priors for $\sigma$ and $r$, in the final analysis of the data; The best-selected models were then fitted using PC priors for the Mat\'{e}rn field parameters in which
\[{\rm P}(\sigma>0.74)=0.05,~~\&~~{\rm P}(r<49)=0.05.\]
\begin{table}[h!]
\begin{center}
\caption{\label{Tab5} Best subset selection for the groundwater data$^\dag$}
\begin{tabular}{cccc}
\hline
Model & WAIC & DIC & Log.score\\
\hline
$ \beta_0 + \beta_1{\rm na} + \beta_2{\rm th} + \beta_3{\rm k} + \beta_4{\rm cl} + \beta_5{\rm hco}_3 + \beta_6{\rm tds} + f(s_i) $ & 18336.51 & 236.823 & 369.880\\
$ \beta_0 + \beta_1{\rm na} + \beta_2{\rm th} + \beta_3{\rm k} + \beta_4{\rm cl} + \beta_6{\rm tds} + f(s_i)$ & 11483.23 & 228.155 & 370.847 \\
$ \beta_0 + \beta_1{\rm na} + \beta_2{\rm th} + \beta_3{\rm k} + \beta_4{\rm cl} + f(s_i)$ & 8650.328 & 223.160 & 348.322\\
$ \beta_0 + \beta_1{\rm na} + \beta_2{\rm th} + f(s_i) $ & {\bf 652.540} & {\bf 221.052} & {\bf 330.339} \\
\hline
\end{tabular}
\end{center}
$^\dag$ \footnotesize{Bold values identify the selected model.}
\end{table}

Table \ref{Tab6} shows the posterior inference for model parameters corresponding to all three models. The estimate of the dispersion parameter $\alpha$ for the GC model describes an overdispersion in the data. This conclusion is in agreement with the classical test of Dean and Lawless (1989). Both covariates (sodium and total hardness) have a significant negative effect on the response; i.e., lower counts of acceptable quality of water for a drink are associated with higher values of both sodium and total hardness. These results are in line with relevant studies regarding the positive association of the chemical elements of groundwater/drinking water with a decline in drinking water quality (Rapant et al. 2017).

Of course, one may notice that other chemical elements should be examined for accurate modeling of these data. We, indeed, considered them implicitly, because:
\begin{enumerate}
\item 
According to the world health organization (WHO) report on "Hardness in Drinking-water" (Cotruvo \textit{et al}., 2011), the principal natural sources of total hardness in water are dissolved polyvalent metallic ions from sedimentary rocks, seepage, and runoff from soils. Calcium and magnesium, the two principal ions, are present in many sedimentary rocks. It would also be important to note that, the most significant relationship between health indicators of human populations and the chemical composition of groundwater was documented as calcium, magnesium, and calcium + magnesium (Rapant et al. 2017). 
\item 
A minor contribution to the total hardness of water is made by other polyvalent ions, such as aluminum, barium, iron, manganese, strontium, and zinc. 
\item 
As mentioned above, adjusting for the sodium and total hardness factors, indirectly, adjust for other factors such as chloride, bicarbonate, and total dissolved solids as well, due to the presence of a high correlation between them (see Figure \ref{fig4}).
\end{enumerate}

\small
\begin{table}[h!]
\begin{footnotesize}
\begin{center}
\caption{\label{Tab6} Posterior means and $95\%$ credible intervals for regression and Mat\'{e}rn field parameters, and the values of WAIC, DIC, and Log.score for groundwater data$^\dag$}
\begin{tabular}{cccc}
\hline
 & GC & Poisson & NB  \\
\hline	
Dispersion & $\alpha=0.309(0.203,0.434)$ & - & ${\rm size}=4.330(1.557,7.757)$  \\
Intercept & $5.778(4.642,6.954) $ & $5.180(4.230,6.093)$ & $5.068(4.141,6.034)$  \\
th & $-0.567(-0.717,-0.425)$ & $-0.398(-0.488,-0.311)$ & $-0.447(-0.561,-0.340)$ \\
na & $-0.010(-0.012,-0.007)$ & $-0.008(-0.010,-0.007)$ & $-0.007(-0.009,-0.005)$ \\ 
$r$ & $64.462(21.636,118.297)$ & $45.028(19.617,76.932)$ & $53.672(21.906,90.656)$ \\
%$\kappa$ & $0.076(0.015,0.158)$ & $0.112(0.035,0.210)$ & $0.092(0.035,0.163)$ \\
%$\tau$ & $6.870(2.312,12.665)$ & $3.730(1.478,6.417)$ & $5.486(2.153,9.409)$  \\
$\sigma^2$ & $0.606(0.073,1.401)$ & $0.795(0.210,1.603)$ & $0.458(0.056,1.083)$ \\ 
WAIC & ${\bf 653.825}$ & $712.799$ & $705.891$ \\
DIC & ${\bf 214.655}$ & $340.357$ & $243.328$ \\
Log.score & ${\bf 330.221}$ & $369.060$ & $354.288$ \\
\hline
\end{tabular}
\end{center}
$^\dag$ \footnotesize{Bold values identify the selected model.}
\end{footnotesize}
\end{table}

\begin{figure}[!htbp]
\begin{center}
\begin{tabular}{ccc}
\includegraphics[scale=0.21]{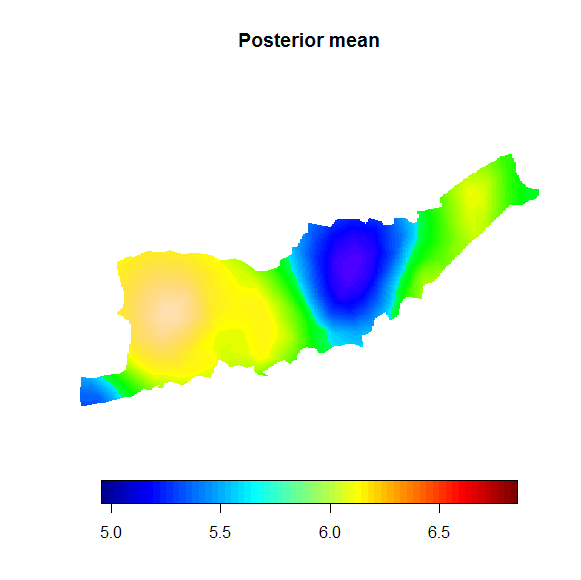}&\includegraphics[scale=0.21]{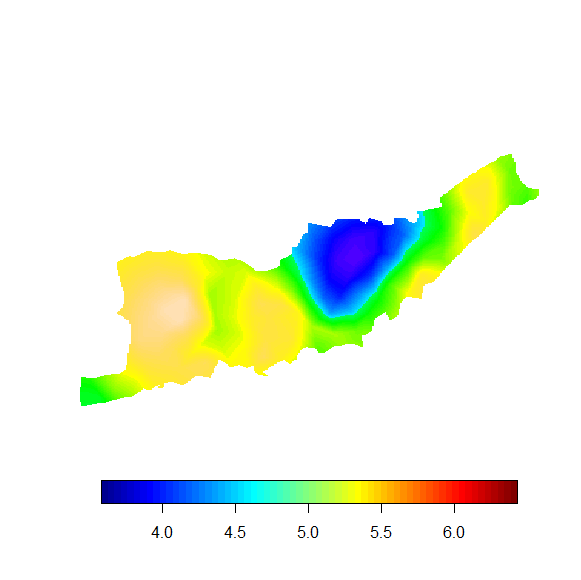} & \includegraphics[scale=0.21]{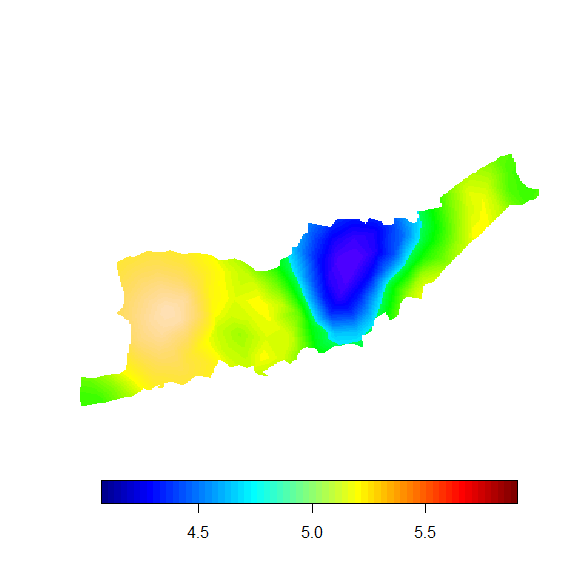}\\
\includegraphics[scale=0.21]{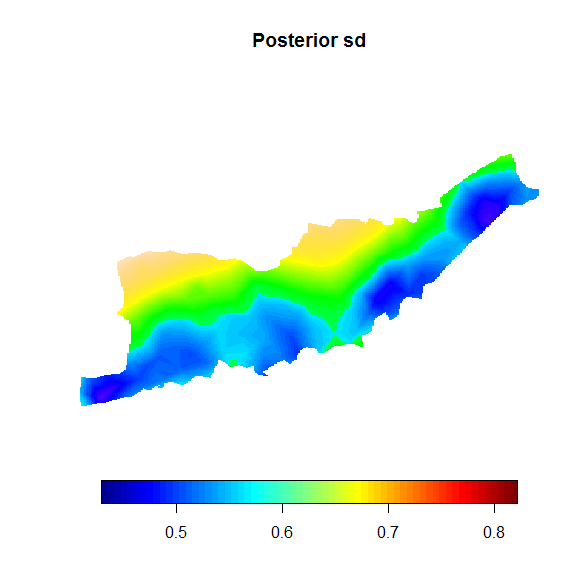}&  \includegraphics[scale=0.21]{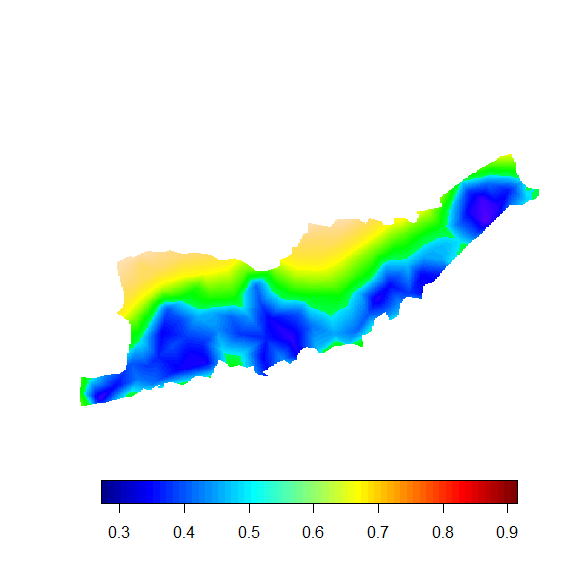} & \includegraphics[scale=0.21]{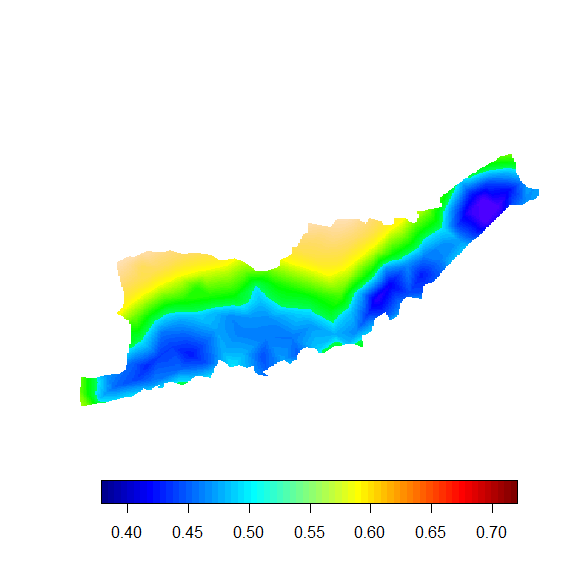}
\end{tabular}
\caption{\label{fig6} \footnotesize{Prediction maps of the response variable (in the log scale) including posterior mean (first row) and posterior sd (second row), obtained from the GC model (first column), Poisson model (second column), and NB model (third column).}}
\end{center}
\end{figure}

The estimates of the marginal variance for the three models are in agreement, especially for the GC and NB models. However, we can see a substantial difference in the practical range of the spatial field. Based on the GC model, with a posterior mean of 64 km for the range, we can assume that the data are characterized by a medium spatial correlation (the maximum distance between coordinates is equal to 195 km). Hence, the estimated spatial effect, under the GC model, is smoother than the corresponding effects based on the Poisson and NB models. This result is also apparent in Figure \ref{fig6}, that shows the posterior mean and standard deviation of the prediction map of the response variable (in the log scale). The smooth predictors in Figure \ref{fig6} display the areas with a better quality of drink water, mostly far from the central parts of the study region. Furthermore, the standard deviation maps reveal that the higher uncertainty parts are located in the northern of the area.

Table \ref{Tab6} also includes the values of WAIC, DIC and logarithmic scores. All three criteria provide strong support for the GC model and indicate that our proposed model is the most appropriate sampling distribution for the counts.

\section{Discussion}\label{Sec6}
Modeling spatial counts with a Poisson regression model is a common practice in many applications. However, count data usually have various levels of dispersion and, consequently, this inherent property of counts should be included in the model. In this paper, we proposed a hierarchical Bayesian approach for modeling spatially dispersed counts by applying the renewal theory that relates nonexponential waiting times between events and the distribution of the counts. Particularly, we extended the methodology based on the gamma distribution assumption for waiting times, which generates a new distribution, named gamma-count, for counts. Our proposed model framework is flexible in terms of allowing various types of dispersions from under-dispersion to over-dispersion. 

Model fitting and inference in a Bayesian spatial GC model can be carried out using MCMC methods, which in this model may come with serious problems regarding convergence and computational time. To overcome these difficulties, we proposed to use a high-speed, non-sampling based, estimation method of Rue et al. (2009) based on INLA. One of the advantages of the INLA approach is that there is a package, called R-INLA, that can be used in the free software R, and consequently, practitioners have the methodology at their disposal.

The results of real example show that our methodology outperforms the traditional models in the scientific applications, a conclusion that is corroborated by the simulation study. The selected model in the groundwater data example reveals relevant findings, agreeing with previous results about the positive association of the chemical elements of drinking water with a decline in drinking water quality. The prediction map shows how the intensity of quality of groundwater changes over the study region in Golestan province. The lowest quality of the water drink could be detected in the central parts of the area, something that may be explained by the potential pollution arising from the use of chemical fertilizers, and the disposal of municipal sewage. All these findings may provide a complete picture of the intensity of groundwater quality and to recognize areas with low quality of groundwater where intervention or prevention programs may be advisable.

An essential concern in the Bayesian inference is the hyperprior distributions, as they can affect the posterior distribution. Here, we chose a gamma prior for the dispersion parameter of the GC model, $\alpha$. It would be interesting to develop a PC prior for $\alpha$ as future work.
Finally, a potential restricting factor in our current spatial GC regression model is that it assumes a constant dispersion level across all observations. For another future work, it would be desirable to model the dispersion parameter, $\alpha$, as a function of covariates and structured effects as well. 

%\newpage

\end{document}